\documentclass[12pt]{article}

\usepackage{epsfig}

\input{epsf}

\begin{document}

\thispagestyle{empty}

\title{Radiation of mixed layer near-inertial oscillations into the
ocean interior}

\author{J. Moehlis$^1$, Stefan G. Llewellyn Smith$^2$}

\maketitle

\noindent
$^1$Department of Physics, University of California, Berkeley, CA 94720\\
$^2$Department of Mechanical and Aerospace Engineering, University of 
California, San Diego \\ 9500 Gilman Drive, La Jolla, CA 92093-0411

\begin{abstract}
\setcounter{page}{1}
The radiation from the mixed layer into the interior of the ocean of near-inertial oscillations excited by a passing storm
in the 
presence of the beta effect is reconsidered as an initial-value
problem.  Making use of the fact that the mixed layer depth is much 
smaller than the total depth of the ocean, the solution is obtained in the 
limit of an ocean that is effectively infinitely deep.  For a
uniform initial condition, analytical results for the velocity,
horizontal kinetic energy density and fluxes are obtained.
The resulting decay of near-inertial mixed layer energy
in the presence of the beta effect occurs on a timescale
similar to that observed.
\end{abstract}

\section{Introduction}

%Horizontal motion of a free particle on the Earth's surface subject only
%to the Coriolis force is governed by the equations
%\[
%\frac{du_p}{dt} - f v_p = 0, \qquad \frac{dv_p}{dt} + f u_p = 0,
%\]
%where $u_p$ and $v_p$ are respectively the eastward and northward components of the 
%particle's velocity in the frame rotating with the Earth,
%$f \equiv 2 \Omega_E \sin \phi$ is the Coriolis parameter, $\Omega_E$ is
%the frequency of the Earth's rotation, and $\phi$ is the latitude
%(see, e.g., \cite{cush94}).  This has
%solution $u_p+iv_p = e^{-i f t} (u_0 + i v_0)$, where $u_0$ and $v_0$ are
%the initial components of the velocity.  This corresponds to the particle's
%velocity describing a circle of radius $(u_0^2 + v_0^2)^{1/2}/f$ with frequency
%$f$.  In the northern hemisphere, $f>0$ and the particle rotates in a
%clockwise direction when viewed from above.
%The inertial frequency $f$ is the low-frequency cutoff for 
%internal waves in the ocean.  An internal wave with frequency near $f$ is 
%called a near-inertial oscillation (NIO).  About half of the total kinetic 
%energy associated with internal waves in the ocean is contained in NIOs
%\cite{dasa85}.

There is much observational evidence, starting with Webster (1968)
and Pollard and Millard (1970), that storms can excite near-inertial 
currents in the mixed layer of the ocean. This phenomenon is evident in observations from the Ocean 
Storms Experiment (D'Asaro et al.\ 1995, Levine and Zervakis 1995,
Qi et al.\ 1995).
Simple models which treat the mixed layer as a solid slab have been quite
successful at explaining the process by which wind generates such currents
(see, e.g., Pollard and Millard (1970), D'Asaro (1985)).
A weakness of the model of Pollard and Millard (1970) is that it explains 
the decay of these currents with an {\it arbitrary} decay constant.  Much 
subsequent work has attempted to determine the detailed characteristics of 
this decay, with possible mechanisms including nonlinear interactions which 
transfer energy to other frequencies (Henyey et al.\ 1986), turbulent 
dissipation (Hebert and Moum 1993), and the radiation of downward propagating 
near-inertial oscillations (NIOs) excited by inertial pumping into the 
interior of the ocean (Gill 1984).  The downward radiation of NIOs will be 
the focus of this paper.  
%Note that about half of the total kinetic energy 
%associated with internal waves in the ocean is contained in NIOs 
%(D'Asaro 1985), and downward propagation of NIOs is believed to be a 
%significant mechanism for mixing in the upper ocean.

Observations give a timescale for the decay of the energy deposited
by the passing storm on the order of ten to twenty days 
(D'Asaro et al.\ 1995, Levine and Zervakis 1995, Qi et al.\ 1995).
This timescale stands in contrast with estimates such as that by 
Gill (1984) that near-inertial currents decaying through the downward
propagation of NIOs and with a horizontal length scale typical of the 
atmospheric forcing mechanism can remain in the mixed layer for longer 
than a year.  To account for this difference, several mechanisms for the enhancement of vertical propagation 
of NIOs have been suggested.  D'Asaro (1989) demonstrated that 
the $\beta$-effect causes a reduction of horizontal scales because the 
meridional wavenumber evolves according to $l = l_0 - \beta t$, where $l_0$ 
is the initial wavenumber, and $l<0$ corresponds to southward
propagation; this accelerates the rate of inertial pumping of energy out of 
the mixed layer, thereby enhancing the decay.  The decay is also enhanced 
through interaction with background geostrophic or quasigeostrophic flow 
(e.g.\ Balmforth et al.\ 1998, Balmforth and Young 1999, and 
van Meurs 1998).

This paper reconsiders the vertical propagation of near-inertial energy 
deposited into the mixed layer by a storm, in the presence of the $\beta$-effect, using a different approach from that of D'Asaro (1989).
The analysis uses the formalism of Young and Ben Jelloul (1997) which is 
outlined in Section 2.  In Section 3, a simplified model with 
three main assumptions is presented.  First, the background flow is assumed 
to be constant in the zonal direction (i.e.\ independent of longitude with zero vorticity).
Second, the buoyancy frequency is taken to be small in 
the mixed layer, and constant in the ocean interior (i.e.\ beneath the mixed layer).
Third, it is assumed that the storm has moved very rapidly across the ocean 
and has created a horizontally uniform near-inertial current to the east 
concentrated within the mixed layer: it is the subsequent evolution of
this motion that is examined.  Section 4 uses the fact that the 
depth of the ocean is very much larger than the mixed layer depth to 
formulate and solve the model for an ocean which is effectively infinitely
deep.  Section 5 discusses the results and suggests directions for further 
investigation.

\section{The NIO equation}
We consider an ocean of infinite horizontal extent and depth
$D$, with the mixed layer comprising the region
$-H_{\rm mix}<z<0$, and the rest of the water column occupying $-D<z<-H_{\rm mix}$.  
The $x$ and $y$ axes are taken to point to the east and north, respectively.
The buoyancy frequency $N = N(z)$ is an arbitrary piecewise 
continuous function of depth~$z$.
  
Young and Ben Jelloul (1997) derive an evolution equation for a 
complex field 
$A(x,y,z,t)$ which governs leading-order NIO motion in the presence of a
steady barotropic background flow and the $\beta$-effect:
\begin{equation}
LA_t + \frac{\partial (\psi,LA)}{\partial (x,y)} + \frac{i}{2} f_0 \nabla^2 A + i \left( \beta y + \frac{1}{2} \zeta \right) LA = 0,
\label{NIO}
\end{equation}
where
\begin{equation}
LA = \frac{\partial}{\partial z} \left( \frac{f_0^2}{N^2} \frac{\partial A}{\partial z} \right),
\label{L}
\end{equation}
$\psi$ is the streamfunction for the background flow, 
$\zeta \equiv \nabla^2 \psi$ is the associated vorticity, and the
Coriolis parameter is $f = f_0 + \beta y$.  Here $\nabla$ is
the horizontal gradient, and $\nabla^2 = \partial_x^2 + \partial_y^2$.
Subscripts denote partial differentiation.
%The asymptotic expansion used in the derivation of equation (\ref{NIO})
%relies upon the frequency of near-inertial waves being close to the inertial
%frequency $f_0$.
The NIO velocity field $(u,v,w)$, buoyancy $b$, and pressure $p$ are
given by
\begin{eqnarray*}
u + i v &=& e^{-i f_0 t} L A, \\
w &=& -\frac{1}{2} f_0^2 N^{-2} (A_{xz} - i A_{yz}) e^{-i f_0 t} + c.c., \\
b &=& \frac{i}{2} f_0 (A_{xz} - i A_{yz}) e^{-i f_0 t} + c.c., \\
p &=& \frac{i}{2} (A_x - i A_y) e^{-i f_0 t} + c.c. 
\end{eqnarray*}
The buoyancy $b$ is related to the density $\rho$ by
\[
\rho = \rho_0 \left[ 1 - \frac{1}{g} \int_0^z N^2(z') dz' - \frac{b}{g} \right],
\]
where $\rho_0$ is the reference density at the top of the ocean.  The pressure
$p$ has been normalized by $\rho_0$.

The boundary conditions are that $A_z=0$ at $z=0$ and $z=-D$.  This
ensures that $w$ vanishes at the top and bottom of the ocean.  Using these
boundary conditions,
\begin{equation}
\int_{-D}^0 (u+i v) = 0.
\label{no_barotropic}
\end{equation}
Thus barotropic motion is not included in the analysis. However
Gill (1984) has shown that the barotropic response to a storm is instantaneous 
and the associated currents are weak.

\section{A Simplified Model}
To simplify the analysis, we assume that $A$ and $\psi$ 
do not vary in the $x$-direction, and that $\zeta=0$.  The analysis thus neglects the effect of background barotropic 
vorticity but crucially
keeps the $\beta$-effect.  The buoyancy frequency profile is taken to be
\begin{eqnarray*}
N^2 &=& \epsilon^2 N_0^2, \qquad -H_{\rm mix} < z < 0, \\
N^2 &=& N_0^2, \qquad -D<z<-H_{\rm mix},
\end{eqnarray*}
where $\epsilon \ll 1$.  Finally, the storm is assumed to have
produced an initial condition of a horizontally uniform near-inertial 
current to the east concentrated within the mixed layer.
Instead of approaching this problem by use of an integral operator as
in
D'Asaro (1989) or by projecting onto normal modes (e.g., Gill 1984, 
Balmforth et al.\ 1998), the problem will be formulated as an initial value 
problem on a semi-infinite domain corresponding to an ocean that is effectively infinitely 
deep.  In order to formulate the problem properly for this limit, this 
section considers an ocean of finite depth.  In Section 4 the solution 
in the limit that the depth of the interior is much greater than the
mixed layer depth will be found.

This formulation as a radiation problem which ignores the presence of
the ocean bottom requires the projection of the initial condition to
be spread across all the normal modes. This is certainly true for small
mixed layer depths in the model of Gill (1984), as shown in Table 1 of
that paper; also see Table 1 of Zervakis and Levin (1995). For deeper 
mixed layers, this is no longer true since
half the initial energy becomes concentrated in the first two or three
modes. However, as pointed in Section 7 of Gill (1984), the depth of
the ocean ``influences the rate of loss of energy by imposing
modulations on the rate, but the average rate of loss is not affected
very much by depth changes''. Hence the results presented here should
be qualitatively relevant even when the continuum assumption is not valid.

\subsection{Nondimensionalization}
Quantities are nondimensionalized according to
%\[
%\hat{y} = \frac{y}{Y}, \qquad \hat{z} = \frac{z}{H_{\rm mix}} + 1, \qquad \hat{t} = \Omega t, \qquad \hat{N} = \frac{N}{N_0},
%\]
\[
\hat{y} = y/Y, \qquad \hat{z} = 1+ z/H_{\rm mix}, \qquad \hat{t} = \Omega t, \qquad \hat{N} = N/N_0,
\]
where
\[
Y \equiv \left( \frac{H_{\rm mix}^2 N_0^2}{\beta f_0} \right)^{1/3}, \qquad \Omega \equiv \left( \frac{\beta^2 H_{\rm mix}^2 N_0^2}{f_0} \right)^{1/3}.
\]
Typical values $\beta = 10^{-11}$ ${\rm m}^{-1} {\rm s}^{-1}$, 
$H_{\rm mix} = 100$ ${\rm m}$, $f_0 = 10^{-4}$ ${\rm s}^{-1}$,
$N_0 = 10^{-2}$ ${\rm s}^{-1}$ give $Y = 10^5$ ${\rm m}$ and
$\Omega = 10^{-6}$ ${\rm s}^{-1}$.  The relevant timescale is thus
$\Omega^{-1} = 11.5$ days.  
Also, the velocity and the field $A$ are nondimensionalized by
\[
(\hat{u},\hat{v}) = \frac{(u,v)}{U}, \qquad \hat{A} = \frac{f_0^2}{U N_0^2 H_{\rm mix}^2} A,
\]
where $U$ is a characteristic value of the initial velocity.

The hats are now dropped for ease of notation.
With this nondimensionalization, 
the buoyancy frequency profile is
\begin{eqnarray*}
N^2 &=& \epsilon^2, \qquad 0<z<1, \\
%N^2 &=& 1 \qquad -H \equiv -\frac{D}{H_{\rm mix}} + 1 < z < 0,
N^2 &=& 1, \qquad -H \equiv 1 -D/H_{\rm mix} < z < 0,
\end{eqnarray*}
and the NIO equation (\ref{NIO}), the boundary conditions, and initial condition become
\begin{eqnarray}
A_{zzt} \!\!\!\!&+&\!\!\!\! \frac{i}{2} N^2 A_{yy} + i y A_{zz} = 0, \label{NIO_nondim} \\
A_z &=& 0, \qquad z = -H, \;\; z = 1, \label{BC} \\
A_{zz} &=& N^2 (u + i v), \qquad t=0. \label{IC} 
\end{eqnarray}
The requirement that $u$ and $v$ remain finite imply the jump conditions
\begin{equation}
A_z|_{z=0^+} = \epsilon^2 A_z|_{z=0^-}, \qquad A_{yy}|_{z=0^+} = A_{yy}|_{z=0^-},
\label{jump_conditions}
\end{equation}
where $z=0^+$ and $z=0^-$ are the limits as $z\rightarrow 0$ from positive
and negative $z$ values, respectively.

This nondimensionalization allows some immediate conclusions to be drawn
about the propagation of NIO energy downwards.  Most importantly, 
if $H_{\rm mix}$ increases, then the timescale $\Omega^{-1}$ decreases.  Thus, 
assuming that the storm causes a uniform near-inertial current throughout
the whole mixed layer, energy transfer will be faster for 
a deeper mixed layer.  This confirms the results of Gill (1984), which
associated the more efficient transfer with a larger projection 
of the initial velocity profile on the first vertical mode.

\subsection{Boundary Condition at the Base of the Mixed Layer}
Expanding $A(y,z,t) = A_0(y,z,t) + \epsilon^2 A_2(y,z,t) + {\cal
O}(\epsilon^4)$ for $0<z<1$,
(\ref{NIO_nondim}) becomes at ${\cal O} (\epsilon^0)$
%\[
%A_{zzt} + \frac{i}{2} \epsilon^2 A_{yy} + i y A_{zz} = 0.
%\]
%Expanding 
%$A(y,z,t) = A_0(y,z,t) + \epsilon^2 A_2(y,z,t) + {\cal O}(\epsilon^4)$,
\[
A_{0zzt} + i y A_{0zz} = 0.
\]
Integrating this subject to the boundary condition that $A_z$ 
and thus $A_{0z}$ vanishes at $z=1$ implies that $A_0$ is independent of $z$.
At ${\cal O}(\epsilon^2)$,
\begin{equation}
A_{2zzt} + i y A_{2zz} + \frac{i}{2} A_{0yy} = 0,
\label{NIO_ML}
\end{equation}
which may be integrated subject to the boundary condition that $A_{2z}$ 
vanishes at $z=1$ to give
\[
A_{2zt} + i y A_{2z} + \frac{i}{2} A_{0yy} (z-1) = 0.
\]
Evaluating at $z=0^+$ and using
$A_{yy} = A_{0yy} + {\cal O}(\epsilon^2)$ and
$A_z = \epsilon^2 A_{2z} + {\cal O}(\epsilon^4)$,
\[
A_{zt} + i y A_z - \frac{i \epsilon^2}{2} A_{yy} = {\cal O}(\epsilon^4), \qquad z=0^+.
\]
Finally, applying (\ref{jump_conditions}) gives the upper boundary 
condition for the NIO field in the ocean interior to leading order in $\epsilon$:
\begin{equation}
A_{zt} + i y A_z - \frac{i}{2} A_{yy} = 0 \qquad z = 0^-.
\label{eff_BC}
\end{equation}
Results obtained in the ocean interior using
(\ref{eff_BC}) are in fact leading-order solutions. We shall continue
to use the notation $A$, even though it is really the leading-order term in the expansion.

\subsection{Initial Condition}
Suppose that in a short time compared with the NIO wave propagation time,
the passing storm induces near-inertial currents in the mixed layer with a 
horizontal scale that is much larger than the one under consideration,
and which can hence be taken to be uniform.
%Also, suppose that the storm causes a large velocity only in the mixed layer.
For simplicity, the initial velocity (consistent with equation 
(\ref{no_barotropic})) is assumed to be piecewise constant with depth:  
\begin{eqnarray*}
(u,v) &=& (1,0) \quad \qquad 0<z<1, \\
&=& (-H^{-1},0), \qquad -H<z<0.
\end{eqnarray*}
The weak flow in the ocean interior is necessary to ensure that the
flow has no barotropic component.
Integrating equation (\ref{IC}) with respect to $z$ and using the boundary 
conditions (\ref{BC}) gives at $t=0$
\begin{eqnarray}
A_z &=& \epsilon^2 (z-1), \qquad 0<z<1, \\
A_z &=& -(z + H)/H, \qquad -H<z<0.
\label{IC2}
\end{eqnarray}

\section{Solution for an Infinitely Deep Ocean}
The total depth of the ocean is typically on the order of a hundred times
the depth of the mixed layer. Thus, the limit of infinite depth is
considered.
The initial condition is taken to be equation (\ref{IC2}) with 
$H \rightarrow \infty$.  The boundary condition for $z \rightarrow -\infty$ 
is taken to be $A_{zz} \rightarrow 0$, corresponding to the near-inertial 
velocities vanishing at infinite depth.
%This limit does not invalidate the use of equation (\ref{NIO})
%which assumed hydrostatic balance and thus holds for the ocean having
%depth much smaller than the horizontal scales.  The ocean still in
%reality has finite depth, but for depths just below the mixed layer it
%is {\it effectively} infinitely deep.  
Of course, this limit excludes the
possibility of reflections off the bottom of the ocean which may be
important.
Finally, the boundary condition for $z=0^-$ given by equation (\ref{eff_BC})
is used.  Hence the problem to be solved for the 
semi-infinite domain $z<0$ becomes
\begin{eqnarray*}
A_{zzt} + \frac{i}{2} A_{yy} + i y A_{zz} = 0, && \qquad z<0, \\
A_{zt} + i y A_z - \frac{i}{2} A_{yy} = 0, && \qquad  z = 0^-, \\
A_{zz} \rightarrow  0, && \qquad z \rightarrow -\infty, \\
A_z = -1, && \qquad t = 0.
\end{eqnarray*}

\subsection{NIO velocity field}
These equations may be solved using Laplace transforms.
Here we present only the major results; further details are given in 
Moehlis (1999). We make the transformations $A(y,z,t) = e^{-i y t}
\tilde{B}(z,T)$, $T \equiv t^3/3$, and $\alpha \equiv (1+i)/2$ and
define the Laplace transform of $\tilde{B}$ by
\begin{equation}
b(z,p) \equiv {\cal L}[\tilde{B}] \equiv \int_0^\infty
\tilde{B}(z,T)e^{-p T} dT.
\end{equation}
Then 
\begin{equation}
b(z,p) = -\frac{1}{\alpha} \; \frac{1}{\sqrt{p} +
\alpha} \; \exp \left( \frac{\alpha z}{\sqrt{p}} \right).
\label{b}
\end{equation}
This Laplace transform and its derivatives with respect to $z$ must be
inverted numerically for the ocean interior ($z<0$).  For the top of the
ocean interior ($z=0^-$) however, they may be obtained in closed form.
For example,
%The NIO field $B(z,t)$ can only be obtained numerically in the ocean
%interior. At the top of the ocean interior however, it may be obtained in
%closed form. For example, The velocity field there is
\begin{equation}
A_{zz}(y,0^-,t) = e^{-i y t} \left[ e^{i t^3/6}  {\rm erfc}
\left(\frac{1+i}{2\sqrt{3}} \; t^{3/2}\right) -1 \right].
\end{equation}

\begin{figure}
\begin{center}
\epsfbox{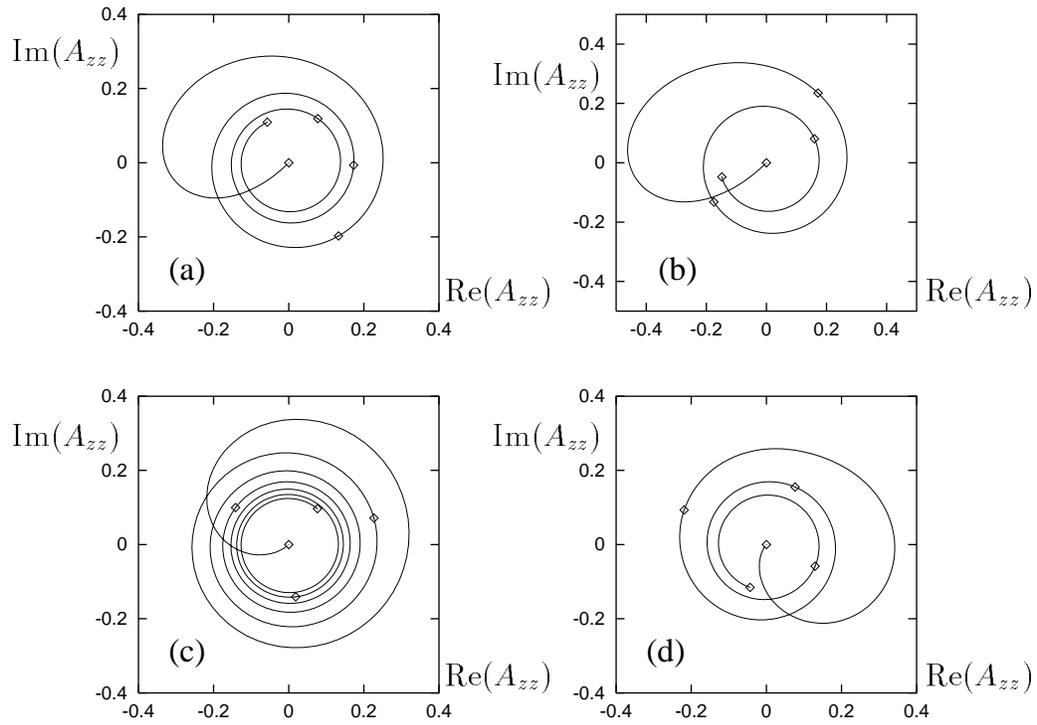}
\end{center}
\caption{Back-rotated velocity for (a) $z=-1$, $y=0$, (b) $z=-0.5$, $y=0$, 
(c) $z=-1$, $y=1$, and (d) $z=-1$, $y=-2$. The diamonds are drawn at
$t=0, 5, 10, 15, 20$.}
\label{hodo_fig}
\end{figure}

We now consider the back-rotated velocity $A_{zz} = e^{i f_0 t}(u + i v)$, 
which filters out purely inertial motion at frequency $f_0$.  
Back-rotated velocities may be represented by hodographs which show the vector 
$({\rm Re}(A_{zz}),{\rm Im}(A_{zz}))$ as curves parametrized by time.  
For $f_0>0$, if these curves are traced out in a clockwise 
(counterclockwise) fashion, the corresponding motion has frequency larger 
(smaller) than $f_0$. Figure \ref{hodo_fig} shows the
back-rotated velocity at different locations. A common characteristic
is that the magnitude of the back-rotated velocity starts at zero,
reaches a peak value shortly after the storm, then decays away.
The depth dependence of the back-rotated 
velocity is seen by comparing Figure \ref{hodo_fig} (a) and (b), 
where both have $y=0$ and thus the same value of the Coriolis parameter $f$.  
Qualitatively the results are the same, but closer to the mixed layer the 
direction change of the back-rotated velocity becomes slower, meaning
that the frequency is closer to $f_0$.
An idea of the latitudinal dependence is seen by comparing
Figure \ref{hodo_fig} (a,c,d): at $y=1$ the hodograph is traced out 
in a clockwise fashion as for $y=0$, but at $y=-2$ it is traced out in a 
counterclockwise fashion.

\subsection{Kinetic energy density and fluxes}
The 
horizontal kinetic energy (HKE) per unit area contained within the mixed layer is
\[
\int_0^1 dz \; \left| \frac{A_{zz}}{N^2} \right|^2 \equiv
\int_0^1 dz \left| \frac{A_{zz}}{\epsilon^2} \right|^2 =
\int_0^1 dz \; |A_{2zz}|^2 .\
\]
Expanding $\tilde{B}(z,T) = \tilde{B}_0(z,T) + \epsilon^2 \tilde{B}_2(z,T) + {\cal O}(\epsilon^4)$
in the mixed layer, (\ref{NIO_ML}) may be used to show that
\begin{equation}
p b_{2zz} - \tilde{B}_{2zz}(z,0) - \frac{i}{2} b_0 = 0,
\label{ML_exact}
\end{equation}
where $b_2 = {\cal L} [\tilde{B}_2]$ and $b_0 = {\cal L} [\tilde{B}_0]$.
The initial condition within the mixed layer is 
$\tilde{B}_{2zz}(z,0) = 1$.
Now $A$ is continuous across $z=0$, and $\tilde{B}_0$ is independent
of $z$ (see Section 3.2). Hence
\[
b_{2zz} = \frac{1}{p} - \frac{i}{2 \alpha p} \frac{1}{\sqrt{p} + \alpha},
\]
which may be inverted to give
\begin{equation}
A_{2zz}(y,t) = e^{-i y t} e^{\alpha^2 t^3/3} {\rm erfc} \left(\frac{\alpha}{\sqrt{3}} \; t^{3/2}\right).
\end{equation}
Therefore the HKE within the mixed layer is
\[
e_{\rm ML} \equiv \left| {\rm erfc}\left( \frac{1+i}{2 \sqrt{3}} \; t^{3/2} \right) \right|^2.
\]

\begin{figure}
\epsfbox{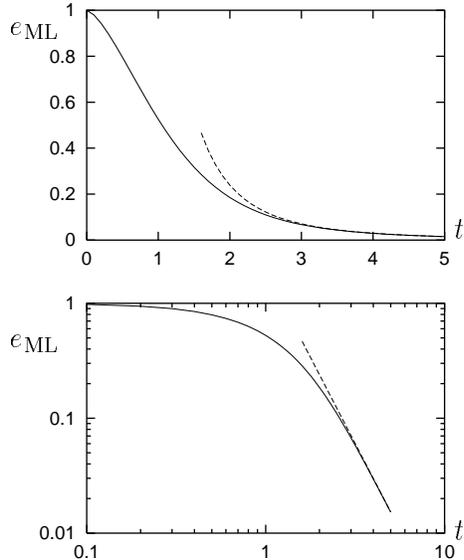}
\caption{ Horizontal kinetic energy per unit  volume (HKE) in the
mixed layer, $e_{\rm ML}$, for linear and logarithmitc axes.
The solid line shows the exact result and the dashed line the
asymptotic result.}
\label{ML_fig}
\end{figure}

The time dependence of $e_{\rm ML}$ is shown in Figure \ref{ML_fig}.
Asymptotic results from Abramowitz and Stegun (1972) for the
complementary error function imply that
\begin{eqnarray*}
e_{\rm ML} &\sim& 1 - \frac{2}{\sqrt{3 \pi}} t^{3/2}, \qquad t \ll 1, \\
e_{\rm ML} &\sim& \frac{6}{\pi t^3}, \qquad t \rightarrow \infty.
\label{energy_ML}
\end{eqnarray*}
Since the energy which leaves the mixed layer enters the interior of the
ocean, this implies that for short times the energy in the interior increases
like $t^{3/2}$.  This does not contradict the result from D'Asaro 
(1989) that for short times the thermocline energy grows like $t^6$.
That result assumes that the wind persists to generate a constant inertially 
oscillating velocity, and that there is no propagating inertial
motion. Here, 
the wind has an instantaneous effect, causing an initial horizontally uniform 
inertial current, and propagating inertial motion is included fully.

%\subsubsection{Energy and Shear Flux}

\begin{figure}[h]
\epsfbox{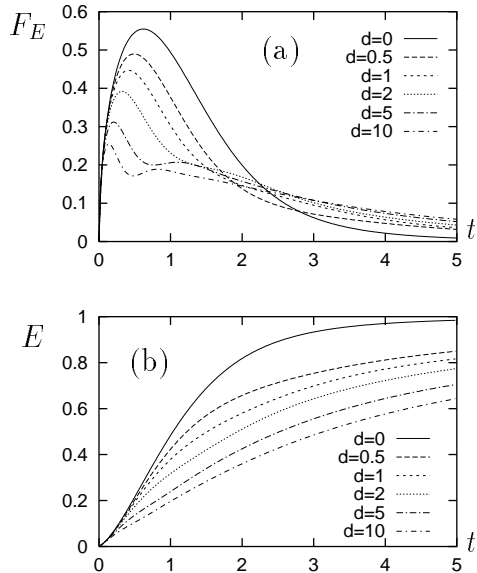}
\caption{(a) $F_E(t;d)$ and (b) $E(t;d)$ for
different depths $d$ below the base of the mixed layer.  These show
instantaneous and time-integrated fluxes of HKE.}
\label{flux_fig}
\end{figure}

Another quantity of interest is the flux of HKE.
Using (\ref{NIO_nondim}) and its complex conjugate gives
\begin{equation}
\frac{\partial}{\partial t} {\rm HKE} = \frac{\partial}{\partial t}\left| \frac{A_{zz}}{N^2} \right|^2 = \frac{i}{2 N^2} \; \frac{\partial}{\partial y} (A_{zz} A_y^* - A_{zz}^* A_y) + \frac{i}{2 N^2} \; \frac{\partial}{\partial z} (A_{yz}^* A_y - A_{yz} A
_y^*).
\label{flux_nondim}
\end{equation}
Assuming $A_{zz} A_y^* - A_{zz}^* A_y$ vanishes for $|y| \rightarrow \infty$
and using equation (\ref{BC}),
\begin{equation}
\frac{d}{dt} \int_{-H}^{-d} dz \int_{-\infty}^{\infty} dx \int_{-\infty}^{\infty} dy \; |A_{zz}|^2 = \int_{-\infty}^\infty \int_{-\infty}^\infty F_E(y,t;d) \,dx \,dy,
\label{intflux_Vd}
\end{equation}
where
\begin{equation}
F_E(y,t;d) \equiv \frac{i}{2} (A_{yz}^* A_y - A_{yz} A_y^*)|_{z=-d}
\label{energy_flux}
\end{equation}
gives the flux of HKE from the region $z>-d$ to the region $z<-d$.  
For this model, we consider the flux per unit area.  
Integrating (\ref{intflux_Vd}) with 
respect to time shows that the quantity
\[
E(t;d) \equiv \int_0^t F_E \,dt
\]
gives the total amount of HKE which has penetrated into the region $z<-d$.
Note that $E(t;d) \rightarrow 1$ corresponds to all the energy
originally in the mixed layer having reached depths below $z=-d$.
Results for $F_E(t;d)$ and $E(t;d)$ obtained by numerically inverting the
appropriate Laplace transforms are shown
in Figure \ref{flux_fig}.  
$F_E$ peaks at the nondimensionalized time 
$t \approx 0.62$; for the typical values quoted in Section 3.1, this 
corresponds to about a week after the storm.  
From Figure \ref{flux_fig}(b) and using the fact that
whatever energy flows through $z=0^-$ must have initally been in the
mixed layer, we see that by $t = 1$ (about 11.5 days after the 
storm) nearly half of the energy associated with horizontal NIO currents
caused by the storm has left the mixed layer; however, only about $38\%$
of the total energy has penetrated below $z=-1$.  By $t=2$ (about 23 days
after the storm), $82\%$ of the
total energy has left the mixed layer, but only $58\%$ has penetrated below
$z=-1$.  Thus, at $t=2$ nearly a quarter of the total energy is
contained in the distance $H_{\rm mix}$ immediately beneath the mixed
layer. This is reminiscent of the accumulation of NIO energy below the
mixed layer seen in Balmforth, Llewellyn Smith and Young (1998).
This model thus gives reasonable estimates for the timescale for which the 
decay of NIO energy occurs: for example, D'Asaro et al.\ (1995) found that 
the mixed layer inertial energy was reduced to background levels by 21 days 
after the storm.

\begin{figure}[h]
\epsfbox{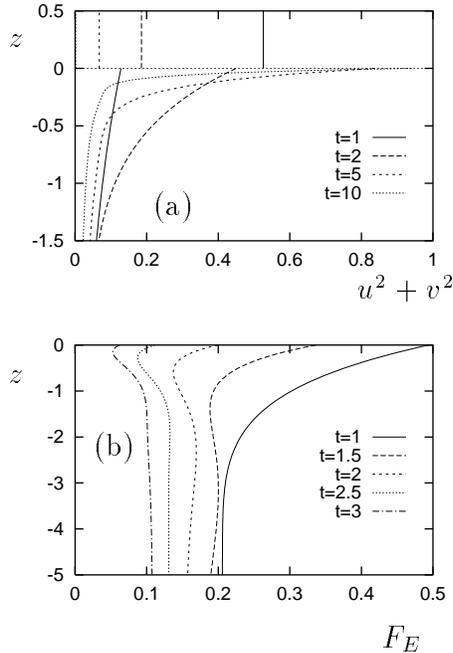}
\caption{Vertical profiles of (a) $u^2 + v^2$ and (b) $F_E(t,|z|)$
at $y=0$ for different times showing the decay of energy 
from the mixed layer ($0<z<1$) and resultant behavior in the interior 
($z<0$).  Note the different vertical scales.}
\label{profile_fig}
\end{figure}

%\subsubsection{Vertical Profiles}
Figure \ref{profile_fig} shows the vertical dependence of the HKE and $F_E$ 
at different times.
As time increases the instantaneous distribution of HKE
becomes more sharply peaked near the base of the mixed layer, but
remains bounded (asymptotically approaching unity) because of energy
conservation.

\subsection{Large-time behavior}
\begin{figure}[h]
\epsfbox{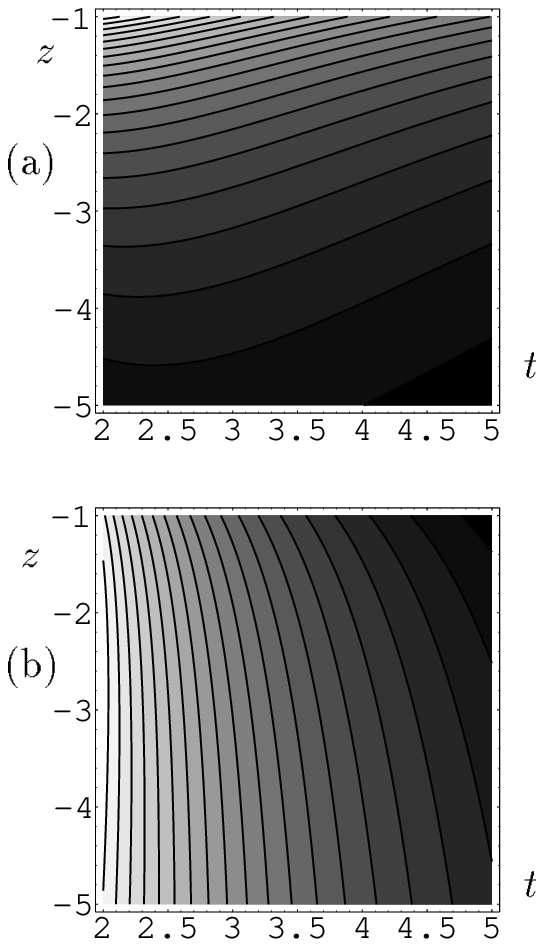}
\caption{Contour plots of the asymptotic results for
(a) $u^2 + v^2$ and
(b) $F_E$.
Darker shading corresponds to smaller values.}
\label{asymptotic_zt_fig}
\end{figure}

The asymptotic behavior of near-inertial properties may be
derived using the method of steepest descents (see Moehlis 1999 for
details).  This shows that in the limit of large $\xi \equiv z^{2/3} t$,
and along the ``rays'' $z=-\eta_0^3 t^3 / 3$,
\[
u^2 + v^2 \sim \frac{2}{(1+\eta_0^2) \pi \eta_0^2 t^3}, \qquad
F_E \sim \frac{2 \eta_0}{\pi (1 + \eta_0^2) t}.
\]
A useful way to represent the asymptotic results is to write $\eta_0$
in terms of $z$ and $t$ and then draw contour plots of quantities of physical
interest in the $(z,t)$ plane: this is shown in Figure \ref{asymptotic_zt_fig}.
In the asymptotic limit for large $\xi$, with $z$ constant,
$u^2 + v^2$ and $F_E$ decrease as time increases.
Note that $\xi$ is large for sufficiently large $z$ and/or $t$.

Finally, Moehlis (1999) also obtained results for the vertical 
shear $u_z^2 + v_z^2$.  To leading order in $\epsilon$, the vertical shear 
within the mixed layer is zero.  The results for vertical shear for the 
interior of the ocean lack physical realism because the model 
allows the shear to grow forever as a consequence of the initial infinite 
shear due to the discontinuity in the initial velocity profile.

\section{Conclusion}
A simplified model has been developed to examine the decay due to
the $\beta$-effect of near-inertial currents excited in the mixed layer by 
a passing 
storm.  This decay occurs due to the radiation of downward propagating
NIOs into the interior of the ocean.  The main assumptions of the
model are that the background flow does not vary in the longitudinal 
direction and has no associated vorticity, that the ocean has a simple (piecewise 
constant) buoyancy frequency profile, and that the storm has moved very quickly 
over the ocean causing a horizontally uniform near-inertial current 
concentrated in the mixed layer.  The $\beta$-effect is included in the 
analysis and is responsible for the radiation of NIOs.  Because the depth of the mixed layer is much smaller than the 
total depth of the ocean, the problem is formulated in the limit of an 
effectively infinitely deep ocean; the resultant initial value problem is 
solved by Laplace transforms.  Analytical results are given for the
horizontal kinetic energy density in the mixed layer, and results from
the numerical inversion of the appropriate Laplace transforms are given
for horizontal kinetic energy, energy flux, and back-rotated velocity.
The asymptotic behavior is also investigated.

Although this simplified model cannot be expected to capture the
full complexity of the aftermath of a storm passing the ocean, it
does capture much of the observed behavior.  Most importantly, in the
presence of the $\beta$-effect the decay of near-inertial mixed layer energy 
is found to occur on the appropriate timescale (approximately twenty days), 
which confirms the analysis of D'Asaro (1989) and observations by 
D'Asaro et al.\ (1995), Levine and Zervakis (1995), and Qi et al.\ (1995).  
The main advantage of the approach described in this paper is that many  
aspects of the decay in the mixed layer are analytically obtained for all 
times, unlike D'Asaro (1989) which predicts the timescale for the decay in a 
short time limit or estimates it in terms of the time it takes normal modes to 
become out of phase (cf. Gill 1984).
Extensions to a more realistic ocean and storm would involve including a
more realistic buoyancy frequency profile (for example, the profile
used by Gill 1984), considering the effect of different initial velocities 
(including both horizontal and vertical structure), and considering the effect
of background flow.  The study of all of these could use the same formalism
of Young and Ben Jelloul (1997) and an approach similar to that presented 
here.

\section*{Acknowledgments}
The majority of this work was carried out at the 1999 Geophysical Fluid
Dynamics program at the Woods Hole Oceanographic Institution.  The authors 
would particularly like to thank W.\ R.\ Young for many useful discussions 
regarding this work.

%\bibliography{gfd1999}

\begin{thebibliography}{99}

%\bibitem{ablo97} Ablowitz, M. J. and Fokas, A. S. (1997) Complex
%Variables: Introduction and Applications, Cambridge University Press,
%647 pp.

\bibitem{abra72} Abramowitz, M. and Stegun, I. A. (1972) Handbook of
Mathematical Functions, Wiley Interscience Publications, 1046 pp.

%\bibitem{arfk95} Arfken, G. B. and Weber, H. H. (1995) Mathematical
%Methods for Physicists (4th ed.), Academic Press, 1029 pp.

\bibitem{balm98} Balmforth, N. J., Llewellyn Smith, S. G. and Young, W. R.
(1998) Enhanced dispersion of near-inertial waves in an idealized
geostrophic flow. {\it J. Mar. Res.}, 56:1--40.

\bibitem{balm99} Balmforth, N. J. and Young, W. R. (1999)  Radiative
damping of near-inertial oscillations in the mixed layer. {\it
J. Mar. Res.}, 57:561--584.

%\bibitem{cush94} Cushman-Roisin, B. (1994) Introduction to Geophysical Fluid
%Dynamics, Prentice Hall, 320 pp.

\bibitem{dasa85} D'Asaro, E. A. (1985) The energy flux from the wind to
near-inertial motions in the surface mixed layer. {\it J. Phys. Oceanogr.},
15:1043--1059.

\bibitem{dasa89} D'Asaro, E. A.  (1989) The decay of wind-forced mixed layer
inertial oscillations due to the $\beta$ effect.  {\it J. Geophys. Res.},
94:2045--2056.

%\bibitem{dasa95a} D'Asaro, E. A. (1995) Upper-ocean inertial currents forced
%by a strong storm.  Part II: Modelling. {\it J. Phys. Oceanogr.},
%25:2937--2952.

\bibitem{dasa95b} D'Asaro, E. A., Eriksen, C. C., Levine, M. D., Niiler, P.,
Paulson, C. A., and van Meurs, P. (1995)  Upper-ocean inertial currents
forced by a strong storm. Part I: Data and comparisons with linear theory.
{\it J. Phys. Oceanogr.}, 25:2909--2936.

%\bibitem{garr79} Garrett, C. and Munk, W. (1979) Internal waves in the
%ocean. {\it Ann. Rev. Fluid Mech.}, 11:339--369.

\bibitem{garr99} Garrett, C.  (1999) What is the ``near-inertial'' band and
why is it different? Unpublished manuscript.

\bibitem{gill84} Gill, A. E.  (1984) On the behavior of internal waves in the
wakes of storms.  {\it J. Phys. Oceanogr.}, 14:1129--1151.

\bibitem{hebe93} Hebert, D. and Moum, J. N.  (1993) Decay of a near-inertial 
wave.  {\it J. Phys. Oceanogr.}, 24:2334--2351.

\bibitem{heny86} Henyey, F. S., Wright, J. A., and Flatt\'{e}, S. M. (1986)
Energy and action flow through the internal wave field: an eikonal
approach. {\it J. Geophys. Res.}, 91:8487--8495.

\bibitem{levi95} Levine, M. D. and Zervakis, V. (1995)  Near-inertial
wave propagation into the pycnocline during ocean storms: observations
and model comparison.  {\it J. Phys. Oceanogr.}, 25:2890--2908.

\bibitem{moeh99} Moehlis, J. (1999)  Effect of a simple storm on a simple
ocean, in {\it Stirring and Mixing, 1999 Summer Study Program in Geophysical
Fluid Dynamics}, Woods Hole Oceanogr. Inst. Unpublished manuscript.

\bibitem{poll70} Pollard, R. T. and Millard, R. C. Jr. (1970) Comparison
between observed and simulated wind-generated inertial oscillations.
{\it Deep-Sea Res.}, 17:813--821.

\bibitem{qi95} Qi, H., De Szoeke, R. A., Paulson, C. A., and Eriksen, C. C.
(1995) The structure of near-inertial waves during ocean storms.
{\it J. Phys. Oceanogr.}, 25:2853--2871.

\bibitem{vanm98} van Meurs, P. (1998) Interactions between near-inertial
mixed layer currents and the mesoscale: the importance of spatial
variabilities in the vorticity field.  {\it J. Phys. Oceanogr.}, 28:1363--1388.

\bibitem{webs68} Webster, F. (1968) Observation of inertial-period motions
in the deep sea. {\it Rev. Geophys.}, 6:473--490.

\bibitem{youn97} Young, W. R. and Ben Jelloul, M. (1997)  Propagation
of near-inertial oscillations through a geostrophic flow. {\it J. Mar. Res.},
55:735--766.

\bibitem{zerv95} Zervakis, V. and Levine, M. D. (1995) Near-inertial
energy propagation from the mixed layer: theoretical considerations.
{\it J. Phys. Oceanogr.}, 25:2872--2889.

\end{thebibliography}

\end{document}